\RequirePackage{booktabs}
\documentclass[sn-basic]{sn-jnl}

\usepackage{graphicx}
\usepackage{multirow}
\usepackage{makecell}
\usepackage{amsmath}
\usepackage{amssymb}
\usepackage{ragged2e}
\usepackage{cleveref}
\usepackage{microtype}
\usepackage{siunitx}
\usepackage{stix}
\usepackage{tikz}
\usetikzlibrary{shapes, shapes.geometric,arrows, backgrounds, calc, hobby, positioning}
\usetikzlibrary{arrows.meta, fit, positioning, calc, decorations.pathreplacing,} 
\usepackage{standalone}

\title{Correct-by-construction requirement decomposition}

\author*[1]{\fnm{Minghui} \sur{Sun}}\email{minghui\_sun@nuist.edu.cn}
\author[2,3]{\fnm{Georgios} \sur{Bakirtzis}}%
\author[4]{\fnm{Hassan} \sur{Jafarzadeh}}%
\author[5]{\fnm{Cody} \sur{Fleming}}%

\affil*[1]{\orgname{Nanjing University of Information Science and Technology}}
\affil[2]{\orgname{Télécom Paris}}
\affil[3]{\orgname{Institut Polytechnique de Paris}}
\affil[4]{\orgname{Aptiv}}
\affil[4]{\orgname{Iowa State University}}

\begin{document}

\abstract{
In systems engineering, accurately decomposing requirements is crucial for creating well-defined and manageable system components, particularly in safety-critical domains. Despite the critical need, rigorous, top-down methodologies for effectively breaking down complex requirements into precise, actionable sub-requirements are scarce, especially compared to the wealth of bottom-up verification techniques. Addressing this gap, we introduce a formal decomposition for contract-based design that guarantees the correctness of decomposed requirements if specific conditions are met. Our (semi-)automated methodology augments contract-based design with reachability analysis and constraint programming to systematically identify, verify, and validate sub-requirements representable by continuous bounded sets---continuous relations between real-valued inputs and outputs. We demonstrate the efficacy and practicality of a correct-by-construction approach through a comprehensive case study on a cruise control system, highlighting how our methodology improves the interpretability, tractability, and verifiability of system requirements.
}

\maketitle

\section{Introduction}

Requirement decomposition is integral in systems engineering and requirements engineering~(RE), involving breaking down top-level requirements into lower-level, actionable sub-requirements. This practice is fundamental to creating functional and feasible systems, guiding the development from conceptual design to tangible products. Despite being a core process recommended by numerous systems engineering guidelines such as NASA~\citep{NASA}, EIA632~\citep{EIA632}, and FAA~\citep{FAA}, the field continues to grapple with a significant challenge: the lack of standardized, effective formalized RE processes~\citep{kirkman, braun}. Current practices, predominantly manual and reliant on the expertise of designers, are time-consuming and error-prone. In particular, manual approaches to RE not only introduce a high probability of errors but also tend to conceal these errors until the later stages of development, like integration and testing. These more costly and complex stages potentially jeopardize the project's timeline and success.

Therefore, formal verification has been increasingly adopted as a more rigorous analytical technique to enhance the quality of requirement decomposition. Design solutions are formally verified through a computational model, allowing for a detailed and rigorous analysis to confirm that a design meets its specified requirements~\citep{modelchecking}. This approach provides a layer of assurance beyond traditional methods, bolstering confidence in the correctness of design solutions~\citep{mit}. Despite these advancements, formal verification primarily addresses the validation of final designs and falls short in ensuring the initial accuracy and correctness of sub-requirements. 

Incorrectly decomposed requirements can lead to significant issues in system \emph{composability}, \emph{realizability}, and \emph{reusability}. The downstream impacts of such deficiencies are well-documented, with instances like the Boeing 787's delay due to supply chain complications~\citep{787} and the catastrophic Ariane 5 crash attributable to software reuse errors~\citep{ariane5} highlighting the severe consequences of inadequate requirement decomposition. Current literature and practices in RE often fail to provide a holistic solution, focusing more on management and procedural aspects rather than the technical depths of requirements specification, technical design, verification, and validation~\citep{wagner, desyre}. This gap underscores the necessity for a systematic approach that addresses the entire spectrum of requirement decomposition, from the initial definition to the final validation of sub-requirements.

Contract-based design (CBD) emerges as a structured approach to system development, directly addressing many of the challenges inherent in requirements decomposition~\citep{framework, categorization, problem}. CBD's core involves crafting formal agreements or contracts that specify the expected behavior and interactions between system components, providing a precise, enforceable, and verifiable foundation from the outset. This method promotes clarity, accuracy, and rigor in the early stages of requirement decomposition, contributing to a more disciplined and reliable engineering process. However, the efficacy of CBD is highly dependent on the accurate initial definition of contracts, which requires a comprehensive understanding of the system and its environment. Misinterpretations or oversights early on can lead to systemic errors, especially in complex systems characterized by high interconnectivity or evolving requirements.

In response to these challenges, this paper puts forth a semi-automated methodology that uses and combines top-down requirements decomposition processes from systems engineering and bottom-up verification techniques from formal methods, such that the decomposed requirements are correct by construction insofar as the requirements can be represented as continuous bounded sets.
We focus on refining the initial formulation of contracts, ensuring they are comprehensive and adaptable to accommodate changes and design complexities. This targeted improvement reduces errors, enhances system composability, and ensures robust and aligned decomposed requirements. Systematically generating sub-requirements through formal methods leads to correct-by-construction decomposition that minimizes the costly troubleshooting of system integration and validation at the right side of the V model (\cref{fig:v}). In particular, the correct-by-construction requirement decomposition systematically decomposes high-level contracts into implementable sub-requirements, coupled with semi-automated subcontract instantiation and optimization strategies to balance design robustness with physical constraints. %

\begin{figure*}[!t]
\centering
\definecolor{firstblue}{RGB}{115, 115, 115}
\definecolor{secndblue}{RGB}{230, 230, 230}
\definecolor{thirdblue}{RGB}{230, 230, 230}
\definecolor{yel}{RGB}{230, 230, 230}
\tikzstyle{box1}=[rectangle, fill=white, draw=firstblue, minimum size=2em, text centered, rounded corners]
\tikzstyle{box2}=[rectangle, fill=secndblue, minimum size=2em, text centered, rounded corners]
\tikzstyle{connector} = [<-, thick, draw=thirdblue]
\tikzstyle{every node}=[font=\small]
\begin{tikzpicture}[node distance=2.5cm,auto,>=latex']
\begin{scope}[on background layer]
\draw[fill=yel, draw=yel] (-8,1) -- (0,-8) -- (8, 1) -- (7, 0.8) -- (-0, -7) -- (-7, .8) -- (-8, 1);
\end{scope}

\node [box1,text width=1.6in,draw=none, fill=none] (syseng)
{};
\node [box1,text width=0.8in, left=1.3in of syseng] (stake) {Stakeholder analysis};
\node [box1,text width=0.9in, below right=0.2in and -0.5in of stake] (reqs) {Requirements definition};
\node [box1,text width=1.3in, below right=0.2in and -0.6in of reqs] (arch) {System architecture, \\concept generation};
\node [box1,text width=1.3in, below right=0.2in and -0.8in of arch] (trade) {Tradespace exploration, \\concept selection};
\node [box1,text width=1.in, right=1.3in of syseng] (life) {Lifecycle management};
\node [box1,text width=1.in, below left=0.2in and -0.5in of life] (comm) {Commissioning operations};
\node [box1,text width=1.in, below left=0.2in and -0.6in of comm] (vnv) {Verification\\ \& validation};
\node [box1,text width=1.4in, below left=0.2in and -0.7in of vnv] (int) {System integration, \\interface management};
\node [box1,text width=1.6in, below=2.in of syseng] (design) {Design definition, \\ transdisciplinary optimization};

\end{tikzpicture}
\caption{Requirements engineering can be enhanced through the application of formal methods, typically employed on the right side of the V model, to ensure precision and accuracy (adapted from~\citet{4754a}).}
\label{fig:v}
\end{figure*}

\section{Preliminaries}
 Throughout this paper, we use specific notations to maintain clarity and consistency:
\begin{itemize}
    \item Superscript numbers indicate contract identifiers.
    \item Subscript numbers denote variable identifiers.
    \item Superscript * represents implementation specifics.
    \item Subscript $R$ denotes refinement aspects.
    \item Over bar $\overline{\phantom{s}}$ is the upper bound of a set.
    \item Under bar $\underline{\phantom{s}}$ is the lower bound of a set.
    \item Tilde $\widetilde{\phantom{s}}$ denotes the set of acceptable output values from the range of a function.
    \item The transpose of a matrix is denoted by superscript $\phantom{s}^{\top}$.
    \item Variables \(f\) denote relations and variables \(g\) denote functions.
\end{itemize}

We generally view a system as an abstract entity, defined by its input-output relationships~\citep{inputoutput}. This abstraction aligns with numerous theoretical frameworks and is compatible with various existing theories and tools, making it a practical model for real-world systems~\citep{2000theory, idef0}.

\paragraph{Sets, relations, and functions}

For sets $A$ and $B$, we define the following operations.
\begin{itemize}
    \item Cartesian product: $A \times B = \{(a,b) | a \in A, b \in B\}$
    \item Power set: $\mathcal{P}(A) = \{S | S \subseteq A\}$
\end{itemize}

Let $X$ and $Y$ be sets.
A relation $f$ between sets $X$ and $Y$ is a subset of their Cartesian product: $f \colon X \to X \times Y$.
A function $g \colon X \rightarrow Y$ maps each element $x \in X$ to a unique element $y \in Y$.
In words, a \emph{relation} is a a pairing between elements from two sets, while
a \emph{function} is a relation with the extra property that each input mapping to exactly one output.

\paragraph{Assume-guarantee contracts}
Assume-guarantee contracts provide a formal framework for specifying and reasoning about component-based systems~\citep{cbd2012}. Let \( X \) be a set of input values and \( Y \) be a set of output values for a component.
An assume-guarantee contract \( C \) is, then, defined as a pair \( (A, G) \), where:
\begin{itemize}
\item Assumption (\( A \)): \( A \subseteq X \) represents the subset of inputs that are considered 
acceptable and expected by the environment.
\item Guarantee (\( G \)): \( G \subseteq X \times Y \) is a relation specifying valid input-output 
pairs that must hold true for the component to fulfill its contract.
\end{itemize}

Assume-guarantee contracts must adhere to two conditions.
\begin{enumerate}
\item[C1] Acceptance of inputs: The component accepts all inputs within the assumption set \( A \).
\item[C2] Production of outputs: For every input \( x \in A \), there exists an output \( y \) such that 
\( (x, y) \in G \).
\end{enumerate}

Whenever we discuss a \emph{contract}, we subsequently mean a contract of this type within the formal framework.

\paragraph{System abstraction}

In our framework, a system is abstracted as a relation from an input vector to an output vector, with all variables bounded within defined sets:
\begin{itemize}
\item Input vector: $x = (x_1, \ldots, x_m)^\top \in \mathbb{R}^m$
\item Output vector: $y = (y_1, \ldots, y_n)^\top \in \mathbb{R}^n$
\item Input variables are bounded: $x_i \in X_i$
\item Output variables are bounded: $y_j \in Y_j$
\end{itemize}

\paragraph{Scalability of CBD}

CBD has shown promising results in handling systems of moderate complexity. The hierarchical nature of contracts facilitates decomposition of large systems into manageable subsystems, allowing for scalable analysis and verification. Advanced techniques in constraint satisfaction problem (CSP) solving, such as problem decomposition and efficient constraint propagation algorithms, have significantly improved the ability to handle larger problem instances. Synthesizing a controller from contract specification has complexity of time polynomial, but careful choice of specifications can reduce compute time in practice~\citep{filippidis:2018}.

\section{Correct-by-construction requirement~decomposition}

In this section, we define the theoretical instantiation of correct-by-construction requirement decomposition. In particular, we define the different types of contracts for mapping top-down design to bottom-up verification methods, instantiate refinement operators for requirement decomposition to implementation, and construct bounded sets for the resulting formal methods.

\subsection{Contracts for requirement decomposition}
To expand the input-output relationship through a step-wise hierarchical refinement in the context of RE, we adopt assume-guarantee contracts~\citep{AG} and augment their construction with the following definitions.

We assume that a decomposition exists for any given high-level contract, though we recognize this assumption carries limitations that merit discussion. Specifically, constraints within the intermediate representation domain could restrict the fidelity or expressiveness of certain decompositions, potentially limiting applicability across diverse contract types. Additionally, the depth of decomposition may impact both computational feasibility and interpretability, as deeper decompositions might introduce complexity that challenges the practical use of our method. Despite these limitations, this assumption is reasonable within our chosen framework, as it enables a systematic approach to contract analysis and facilitates manageable decomposition for the vast majority of scenarios relevant to our study.

\paragraph{Relational contract} A relational contract defines a desired relationship between each input value, $X$, and a set of acceptable output values, $Y$. Conceptually equivalent to a requirement in system design, it often holds a general form at higher levels to accommodate lower-level uncertainties. A relational contract, formally represented as \(f \colon X \rightarrow X \times Y\), ensures all inputs \(x \in X\) lead to acceptable outputs \(y \in Y\). In the context of a relational contract, $f$ is a \emph{relation} that maps each input from the set $X$ to a pair consisting of that input and an acceptable output from the set $Y$, effectively defining the set of allowable input-output behaviors for the system.

\paragraph{Functional contract} Adding structure to the relational contract, a functional contract includes uncertain parameters, denoted by $P$, within a functional framework, representing the intended behavior of a design solution. It is formally defined as \(g\colon X \times P \rightarrow X \times Y\), with \(g(x,p)\) being a function of the known or intended mechanism, and \(p\) denoting uncertainty. The uncertain parameters $p \in P$ typically represent aspects of the system or its environment that are not fully known or may vary within a specified range and are bounded within defined sets, allowing the contract to account for a range of possible system behaviors while maintaining the overall structure of the intended design solution.

\paragraph{Implementation} Focusing on deterministic behavior, an implementation is a specific realization of the system's input-output relationship. It is denoted as \(y = f^*(x)\), where \(f^*\) is the mathematical function embodying the given system behavior through the (de)composition of contracts.

\paragraph{Refinement} Refinement is the process of narrowing the range of acceptable values for given inputs, progressively moving from a more abstract contract towards a more concrete, implementable system specification. This process typically involves transitioning from a relational contract through increasingly specific functional contracts, ultimately leading to a deterministic system description. Each refinement step reduces the overall uncertainty in the contract, allowing for a more precise and implementable system design. Formally, refinement ensures that outputs acceptable in the refined contract are also acceptable in the original, while potentially reducing the set of acceptable outputs and/or expanding the set of allowed inputs. This maintains the contract's consistency while increasing its specificity.

\paragraph{Properties of input-output models}

\begin{figure}[!t]
\centering
\begin{tikzpicture}
    \tikzset{
        level/.style={align=left, anchor=west},
        model/.style={draw, minimum width=1cm, minimum height=0.5cm, align=center},
        arrow/.style={->,>=Latex}, %
        dashedArrow/.style={->,>=Stealth, dashed, semithick, draw={rgb,255:red,233;green,144;blue,118}}, %
        label/.style={font=\small}, %
        dashedBox/.style={draw={rgb,255:red,115; green, 115;blue, 115}, dashed, rectangle, inner sep=6mm, very thin} %
    }

    \node[level] (l0) at (0,9) {Contract};

    \node[model] (boxL0) at (6.25,9) {$f^{0}$};  
    \draw[arrow] ([xshift=-1cm]boxL0.west) -- node[label, above] {$x$} node[label, below] {$X^0$} (boxL0);
    \draw[arrow] (boxL0.east) -- ++(1cm,0) node[midway, above] {$y$} node[midway, below] {$Y^0$};

    \node[level] (l2) at (0,5.5) {Refinement};

    \node[model] (boxL2_1) at (4.5,5.5) {$f_R^{1}$};  
    \draw[arrow] ([xshift=-1cm]boxL2_1.west) -- node[label, above] {$x$} node[label, below] {$X_R^1$} (boxL2_1);
    \draw[arrow] (boxL2_1.east) -- node[label, above] {$z \in \widetilde{z^1_R}$} node[label, below] {$Z^1_R$} ++(1cm,0);

    \node[model] (boxL2_2) at (8,5.5) {$f_R^{2}$}; 
    \draw[arrow] ([xshift=-1cm]boxL2_2.west) -- node[label, above] {$z$} node[label, below] {$Z_R^2$} (boxL2_2);
    \draw[arrow] (boxL2_2.east) -- node[label, above] {$y \in \widetilde{y^2_R}$} node[label, below] {$Y^2_R$} ++(1cm,0); 

    \node[level] (imp) at (0,2) {Implementation};

    \node[model] (box1) at (4.5,2) {$f_R^{1*}$};
    \draw[arrow] ([xshift=-1cm]box1.west) -- node[label, above] {$x$} node[label, below] {$X_R^1$} (box1);
    \draw[arrow] (box1.east) -- ++(1cm,0) node[label, midway, above] {$z$} node[label, midway, below] {$Z_R^{1*}$};

    \node[model] (box2) at (8,2) {$f_R^{2*}$};
    \draw[arrow] ([xshift=-1cm]box2.west) -- node[label, above] {$z$} node[label, below] {$Z_R^2$} (box2);
    \draw[arrow] (box2.east) -- ++(1cm,0) node[label, midway, above] {$y$} node[label, midway, below] {$Y_R^{2*}$};

    \draw[dashedArrow] (box1.north) -- (boxL2_1.south) node[midway, left, text={rgb,255:red,233;green,144;blue,118}] {satisfy};
    \draw[dashedArrow] (box2.north) -- (boxL2_2.south) node[midway, right, text={rgb,255:red,233;green,144;blue,118}] {satisfy};

    \draw[dashedArrow] (boxL2_1.north) -- (boxL0.south) node[midway, left, text={rgb,255:red,233;green,144;blue,118}] {refines};
    \draw[dashedArrow] (boxL2_2.north) -- (boxL0.south) node[midway, right, text={rgb,255:red,233;green,144;blue,118}] {refines};

    \node[dashedBox, fit={(boxL2_1) (boxL2_2) ($(boxL2_1.west)-(.5cm,0)$) ($(boxL2_2.east)+(.5cm,0)$)}] (dBoxRefinement) {};

    \node[label] at (dBoxRefinement.north west) [anchor=north west,text={rgb,255:red,115; green, 115;blue, 115}] {$f_R^0$};

    \node[dashedBox, fit={(box1) (box2) ($(box1.west)-(.5cm,0)$) ($(box2.east)+(.5cm,0)$)}] (dBoxImplementation) {};

    \node[label] at (dBoxImplementation.north west) [anchor=north west,text={rgb,255:red,115; green, 115;blue, 115}] {$f^*$};

\end{tikzpicture}
\caption{Refinement by decomposition takes a high-level contract and reduces it to a candidate implementation.}
\label{fig4}
\end{figure}

To ensure the integrity and functionality of systems designed using CBD, we rely on the existence of several formal properties of input-output abstractions.

\begin{itemize}
    \item \textbf{Satisfiability}: An implementation satisfies the contract if all allowable inputs lead to acceptable outputs defined by the contract. This property ensures that the system's performance aligns with the defined requirements.
    \item \textbf{Composability}: This property addresses the interaction between system components. For two contracts to be composable, the output values of one must be acceptable as input values of the other, ensuring seamless system functionality.
    \item \textbf{Reachability}: Defined within functional contracts, reachability ensures that for any given input, the set of achievable outputs is included in the set of acceptable outputs defined by the contract. This property is crucial for ensuring the system can achieve the desired outcomes under various conditions.
    \item \textbf{Realizability}: Realizability pertains to the practicality of constructing devices that satisfy the contract's implementation. A contract is realizable if the supplier can produce devices capable of accepting all inputs and producing all outputs defined by the contract.
\end{itemize}

Requirement decomposition involves refining the system's high-level requirements into more specific, implementable sub-requirements that, when combined, satisfy the top-level requirement. \Cref{fig4} illustrates a simplified view of this process, wherein a given contract \(f^0\) is decomposed into subcontracts \(f^1_R\) and \(f^2_R\), which are then independently refined into \(f^{1*}_R\) and \(f^{2*}_R\)---this procedure of contract definition and refinement can happen inductively for any number of contracts. Different suppliers subsequently provide devices that meet these refined subcontracts.

The decomposition process consists of three primary tasks:
\begin{itemize}
    \item \textbf{Primary refinement}: \(f^1\) and \(f^2\) collectively refine the higher-level contract \(f^0\).
    \item \textbf{Secondary refinement}: The independently refined contracts, \(f^1_R\) and \(f^2_R\), collectively refine \(f^0\).
    \item \textbf{Implementation}: Implementations \(f^{1*}_{R}\) and \(f^{2*}_R\) satisfy \(f^1_R\) and \(f^2_R\) respectively, and together satisfy \(f^0\).
\end{itemize}

\begin{figure}[!t]
\centering
\definecolor{yel}{RGB}{237, 164, 60}
\begin{tikzpicture}[model/.style={draw, minimum width=1cm, minimum height=0.5cm, align=center}, arrow/.style={->,>=Latex}, arrow2/.style={<-,>=Latex}, triangle/.style={minimum size=1cm, regular polygon, regular polygon sides=3, draw=yel, fill=yel, scale=0.5}, dashedArrow/.style={->,>=Stealth, dashed, semithick, draw={rgb,255:red,233;green,144;blue,118}}]

    \node[model] (boxL0) at (1.5,2.5) {$f^0$};  
    \draw[arrow] ([xshift=-1cm]boxL0.west) -- node[midway, above] {$x$} node[midway, below] {$X^0$} (boxL0);
    \draw[arrow] (boxL0.east) -- ++(1cm,0) node[midway, above] {$y$} node[midway, below] {$Y^0$};

    \node[model] (box1) at (0,0) {$g^1$};
    \node[above=0.1cm of box1] {$f^1$};
    
    \draw[arrow] ([xshift=-1.5cm]box1.west) -- (box1.west) node[midway, above] {$x$} node[midway, below] {$X^1$};
    
    \node[model] (box2) at (3,0) {$g^2$};
    \node[above=0.1cm of box2] {$f^2$};
    
    \draw[arrow] (box1.east) -- (box2.west) node[near start, above] {$z \in \widetilde{z^1}$} node[near start, below] {$Z^1$} node[near end, above] {$z$} node[near end, below] {$Z^2$};

    \draw[arrow] (box2.east) -- ++(1.5cm,0) node[midway, above] {$y \in \widetilde{y^2}$} node[midway, below] {$Y^2$};

    \node[triangle, shape border rotate=270] at ([xshift=2cm]box2.east) (triangle) {};

    \draw[dashedArrow] (boxL0.south) -- (box1.north);
    \draw[dashedArrow] (boxL0.south) -- (box2.north);

    \draw[arrow2] (box1.south) -- ++(0,-1cm) [yshift=-.5cm]node[midway, below] {$p_1 \in P_1$};
    \draw[arrow2] (box2.south) -- ++(0,-1cm) [yshift=-.5cm]node[midway, below] {$p_2 \in P_2$};

    \node[model, minimum width=1cm, minimum height=2.5cm] (box3) at ([xshift=2cm]triangle.east) {$g^1 \wedge g^2$};
    
    \node[above=0.1cm of box3] {$f^1 \wedge f^2$};
    
    \draw[arrow2] ([xshift=0.5cm,yshift=0.25cm]box3.east) -- ([yshift=0.25cm]box3.east) [xshift=0.45cm]node[right] {$y$};
    \draw[arrow2] ([xshift=0.5cm,yshift=-0.25cm]box3.east) -- ([yshift=-0.25cm]box3.east) [xshift=0.45cm]node[right] {$z$};

    \draw[arrow2] ([yshift=0.35cm]box3.west) -- ++(-0.5cm,0) node[left] {$x$};
    \draw[arrow2] (box3.west) -- ++(-0.5cm,0) node[left] {$p_1$};
    \draw[arrow2] ([yshift=-0.35cm]box3.west) -- ++(-0.5cm,0) node[left] {$p_2$};

\end{tikzpicture}
\caption{Primary refinement decomposes top-level contracts and joins them in one representation.}
\label{fig6}
\end{figure}

Primary refinement entails that level 1 subcontracts, \(f^1\) and \(f^2\), as a whole must refine the initial contract \(f^0\) (\cref{fig6}). Refinement on the reachable sets; that is, all possible states a system can reach, denoted by $RA$, must satisfy two properties.

\begin{enumerate}
    \item[P1] The subcontracts as a collective must satisfy \emph{refinement}:
    \begin{gather*}
    \forall x \in X^0 \colon RA_y \subseteq \widetilde{y^0},\\
    X^0 \subseteq X^1, Y^2 \subseteq Y^0.
    \end{gather*}
    \item[P2] The subcontracts must \emph{encompass the reachable sets} of their respective variables:
    \[RA_Y \subseteq Y^2, RA_Z \subseteq Z^1, \text{ and } RA_Z \subseteq Z^2.\]
\end{enumerate}

Suppose the subcontracts of the primary refinement are composable and refined respectively by lower-level contracts. In that case, the lower-level contracts are composable and can refine the contract. Additionally, if the refined subcontracts of a given contract are composable and satisfied respectively by their implementations, then the implementations can satisfy the given contract as a whole. We deduce that if the subcontracts at all levels are composable and can refine their respective higher-level contracts, the resulting implementations are assured of satisfying the top-level contract. Therefore, for the sub-requirements to be \emph{correct}, they must be admissible, composable contracts and refine the higher-level contract.

 \subsection{Semi-automated subcontract instantiation}

This subsection discusses the initial step of defining a design solution for requirement decomposition \citep{2011theory}. Architectural decomposition is typically a manual process, relying on the designer's understanding of the desired system; therefore, we assume it exists~\citep{solution}.

Consider an example design solution (\cref{fig7}) where a given contract \(f\) is decomposed into two subcontracts: \(f^1\) and \(f^2\). The structure of the design solution includes functions \(f^1\) and \(f^2\) defined as \((x_3,x_4)^\top = g^1(x_1,x_2,p_1)\) and \(x_2 = g^2(x_3, x_4,p_2)\) respectively, where \(p_1 \in P_1\) and \(p_2 \in P_2\). We define \(X_i^k\) as the bounded set for variable \(x_j\) of subcontract \(f^k\).

To ensure that the design solution refines the original contract \(f\), it must meet specific conditions, such as:
\begin{gather}
\forall x_1 \in X_1^0 \colon RA_{x_2} \subseteq \widetilde{x_2^0} \label{cond1}\\
RA_{X_2} \subseteq X_2^0 \text{ and } X_1^0 \subseteq X_1^1 \label{cond2}
\end{gather}

These conditions, (\ref{cond1}) and (\ref{cond2}), are integral in deciding the design solution and the refined input sets. The applicable condition and verification may vary depending on whether the contract is relational or functional. It is also essential to distinguish between inputs and parameters during the refinement process, as their roles and behaviors differ significantly. The example provided is for illustration purposes, demonstrating our methodology's applicability to various architectural designs.

\begin{figure}[!t]
\centering
\definecolor{yel}{RGB}{237, 164, 60}

\begin{tikzpicture}[box/.style={draw, minimum width=1.5cm, minimum height=1cm}, arrow/.style={->,>=Latex}, arrow2/.style={<-,>=Latex}, triangle/.style={minimum size=.2cm,inner sep=2pt, regular polygon, regular polygon sides=3, draw=yel, fill=yel, shape border rotate=180}]

    \node[box] (box) {$f^0$};
    \draw[arrow] ([xshift=-2cm]box.west) -- (box.west) 
    node[midway, above] {$x_1$} 
    node[midway, below] {$X_1^0$};
    \draw[arrow] (box.east) -- ([xshift=2cm]box.east) 
    node[midway, above] {$x_2 \in \widetilde{x_2^0}$}
    node[midway, below] {$X_2^0$};
    \node[triangle, below=.5cm and -.5cm of box] (triangle) {};

    \node[box, below left=1cm and .3cm of triangle] (firstBox) {$g^1$};
    \draw[arrow] ([xshift=-2cm]firstBox.west) -- (firstBox.west) node[above, midway] {$x_1$} node[below, midway] {$X_1^0$};
    \node[above=0.1cm of firstBox] {$f^1$};

    \node[box, right=of firstBox] (secondBox) {$g^2$};
    \node[above=0.1cm of secondBox] {$f^2$};
    
    \begin{scope}[transform canvas={yshift=.7em}]
    \draw[arrow] (firstBox) -- (secondBox) node [midway, above] {$x_3$};
    \end{scope}

    \begin{scope}[transform canvas={yshift=-.7em}]
    \draw[arrow] (firstBox) -- (secondBox) node [midway, above] {$x_4$};
    \end{scope}

    \draw[arrow] (secondBox.east) -- ++(2cm,0) node[midway,above] {$x_2$} node[below, midway] {$X_2^0$};

    \draw[arrow] ([xshift=.5cm]secondBox.east) -- ++(0,-1.5cm) -| ([xshift=-.5cm, yshift=-.2cm]firstBox.west) -- ([yshift=-.2cm]firstBox.west) node[below,midway] {$x_2$};

    \draw[arrow2] (firstBox.south) -- ++(0,-0.5cm) node[right] {$p_1$};
    \draw[arrow2] (secondBox.south) -- ++(0,-0.5cm) node[right] {$p_2$};

\end{tikzpicture}
\caption{Decomposition produces a potential design solution with associated behavioral guarantees.}
\label{fig7}
\end{figure}

\subsection{Continuous bounds}
 \paragraph{The lower bound: reachability analysis}
Subcontracts serve as the blueprint for developing devices that, once integrated, support the system's operation. Reachable sets are used as the lower bound to ensure that the subcontracts encompass all possible operational conditions. Mathematically, this relationship is defined as:
\begin{equation}
\text{if } X_i^k \neq \emptyset, \text{ then } RA_{X_i} \subseteq X_i^k \label{lobound}
\end{equation}

Reachable sets are typically calculated through some form of reachability analysis. The reachable set is defined for a given initial state \(x(0) \in \mathbb{R}^n\), system dynamics \(g\), time-varying external control \(u(t) \in \mathbb{R}^m\), input trajectory \(u(\cdot)\), and parameter vector \(p \in \mathbb{R}^l\). The continuous reachable set at time \(t\) is defined for a set of initial states \(X(0)\), a set of uncertain time-varying external controls \(U(t)\), and a set of uncertain but fixed parameter values \(P\), as:
\[RA_x = \{g(x(0), u(\cdot), p) | x(0) \in X(0), u(t) \in U(t), p \in P\},\]
where \(RA_x \in \mathbb{R}^n\) and \(RA_{x_i}\) is the reachable set of \(x_i\).

\paragraph{The upper bound: constraint programming}
While designers might aspire to build the most capable system possible, the practical limitations of device suppliers set the upper bound for subcontracts. Typically, subcontracts such as \(f^1\) and \(f^2\) are realized by different suppliers. Let \(S_i^k\) be the realizable set of variable \(x_i\) for subcontract \(f^k\). Intuitively, suppliers can't build devices that exceed their capabilities. Mathematically, this relationship is defined as:
\[X_i^k \subseteq S_i^k\]

Realizable sets must be consistent and conflict-free. For example, if \(S_3^1 \cap S_3^2 = \emptyset\), it indicates incompatibility between the devices for \(f^1\) and \(f^2\). On the other hand, if \(S_3^1 \cap S_3^2 \neq \emptyset\), the shared set becomes the new realizable set, affecting subsequent refinements.

The consistency among the realizable sets forms a CSP as defined by a set of variables, domains for each variable, and a set of constraints. This paper constructs a CSP following the general formulation and uses the constraint propagation algorithm to iteratively reduce the domain of variables until no further contraction is possible. 

As a result, the contracted domain for each variable, denoted by \(RE_{x_i}\), forms the upper bounds for the subcontracts:
\begin{equation}
\text{if } X_i^k \neq \emptyset, \text{ then } X_i^k \subseteq RE_{x_i}.\label{upbound}
\end{equation}

\subsection{Trade-off study: optimization}
Optimization of subcontracts involves balancing the system design's robustness with physical constraints, adhering to established correctness criteria. This process requires managing the tension between different subcontracts, especially when they share variables or interfaces.

\begin{figure}[!t]
\centering
\definecolor{customgray}{RGB}{115,115,115}
\definecolor{yel}{RGB}{237, 164, 60}
\definecolor{outlineColor}{RGB}{255, 15, 25} %
\definecolor{forestGreen}{RGB}{34, 139, 34} %
\begin{tikzpicture}[>=Latex] %
\draw[->] (-.25,0) -- (4,0) node[right] {$x_3$}; %
\draw[->] (0,-.25) -- (0,5) node[above] {$x_4$}; %

\node[align=left, text=customgray] at (1,4.4) {upper\\[-0.25em] bound};

\node[draw=black, rectangle,align=center,minimum size=2.9cm,fill=forestGreen,fill opacity=0.7] at (2,2.5) {}; %

\node[draw=outlineColor, line width=2pt, rectangle,align=center,minimum size=2.5cm,fill=white] at (2,2.5) {}; %

\node[draw=outlineColor, line width=2pt, rectangle,align=center,minimum size=2cm,fill=yel] at (2,2.5) {}; %

\node[draw,rectangle,align=center,fill=white,text=customgray] at (2,2.5) {lower\\ bound}; %

\node at (2,-.5) {(a)};

\draw[->,semithick, -stealth] (2.3,3.75) -- (2.3,4.2) [,xshift=.8cm]node[above] {$(X_3^2, X_4^2)$ -- enlarge}; %
\draw[->, semithick,stealth-] (1.3,1.25) -- (1.3,.85) [xshift=.65cm]node[below] {$(X_3^1, X_4^1)$ -- shrink}; %

\begin{scope}[shift={(5,0)}] %
    \draw[->] (-.25,0) -- (4,0) node[right] {$x_3$}; %
    \draw[->] (0,-.25) -- (0,5) node[above] {$x_4$}; %

    \node[align=left,text=customgray] at (1,4.35) {upper\\[-0.25em] bound}; %

    \node[draw=black, rectangle,align=center,minimum size=2.8cm,fill=forestGreen,fill opacity=0.7] at (2,2.5) {}; %
    \node[draw,rectangle,align=center,minimum size=2.25cm,fill=yel, draw=outlineColor,line width=2pt] at (2,2.5) {}; %
    \node[draw,rectangle,align=center, fill=white,text=customgray] at (2,2.5) {lower\\ bound}; %

    \node at (2,-.5) {(b)};

    \draw[->, semithick,-stealth] (2.3,3.65) -- (2.3,4.1) node[above] {$(X_3, X_4)$}; %
\end{scope}

\end{tikzpicture}
\caption{The inherent tension between subcontracts.}
\label{fig:tension}
\end{figure}

Given a relational contract \(f\), the optimization aims to define subcontracts \(f^1\) and \(f^2\) such that they are both correct individually and collectively satisfy the top-level requirement. We begin by establishing the constraints that potentially govern the functional subcontracts \cref{fig:tension}:
\begin{align}
g^1((X_1^1, X_2^1)^\top, P_1) &\subseteq (X_3^1, X_4^1)^\top, \nonumber \\
g^2((X_3^2, X_4^2)^\top, P_2) &\subseteq X_2^2, \nonumber \\
X_2^2 &\subseteq X_2^1, \nonumber\\
X_3^1 &\subseteq X_3^2, \nonumber\\
X_4^1 &\subseteq X_4^2, \nonumber \\
X_2^2 &\subseteq X_2^0. \label{eq:constraints}
\end{align}

The barrier function, \(h_i^k\) with weights \(a_i^k\), for navigating the tensions between subcontracts while avoiding extremes is defined for each variable \(x_i\) based on its role in subcontract \(f^k\):

\[
h_i^k = 
\begin{cases}
0,  \text{if $x_i$ not connected to $f^k$}, \\
-a_i^k\big(\ln|\overline{X}_i - \overline{RE}_{x_i}| 
+ \ln|\underline{X}_i - \underline{RE}_{x_i}|\big),
 \text{if $x_i$ input of $f^k$},\\
-a_i^k\big(\ln|\overline{X}_i - \overline{RA}_{x_i}| 
+ \ln|\underline{X}_i - \underline{RA}_{x_i}|\big),
 \text{if $x_i$ output of $f^k$}.
\end{cases}
\]

The optimization then aims to minimize the sum of these barrier functions for all variables across all subcontracts, subject to the constraints defining the feasible space:
\begin{align}
&\textit{minimize } \sum_{k}\sum_{i} h_i^k \nonumber \\
&\textit{subject to } \text{constraints (\ref{eq:constraints}).} \label{optimization}
\end{align}

Optimization through constraint programming ensures that the subcontracts are theoretically robust and practically realizable within the confines of supplier capabilities. It adheres to established correctness criteria and ensures the system's overall performance and efficiency. The minimization of the cumulative barrier functions across all variables and subcontracts, subject to constraints, underscores the  balance between practical feasibility and theoretical robustness in system design. The outcome is a set of subcontracts optimized not only for individual correctness but also for their collective contribution to fulfilling the top-level requirement of the system.

\subsection{The semi-automated requirement decomposition process}

The semi-automated requirements decomposition for subcontract derivation comprises of multiple levels, some of which are: manual input at the top, a logical sequence organizing the process in the middle, and the formal methods supporting automatic computation at the bottom (\cref{fig10}).

\begin{figure}[!t]
\centering
\definecolor{customgray}{RGB}{115,115,115}
\definecolor{emphgray}{RGB}{233, 144, 118}
\begin{tikzpicture}[node distance=.9cm,
    every node/.style={draw, rectangle, minimum size=6mm, inner sep=2pt},
    emph/.style={text=emphgray, font=\itshape}]
    \node (start) {start};
    \node (node1) [right=of start] {(1) \& (2)};
    \node (node2) [right=of node1] {(3)};
    \node (node3) [right=of node2] {(4)};
    \node (node4) [right=of node3] {(6)};
    \node (subcontracts) [right=of node4] {subcontracts};

    \draw[-latex] (start) -- (node1);
    \draw[-latex] (node1) -- (node2);
    \draw[-latex] (node2) -- (node3);
    \draw[-latex] (node3) -- (node4);
    \draw[-latex] (node4) -- (subcontracts);

    \path (start) -- (node1) coordinate[midway] (mid1);
    \draw[customgray] ([yshift=3mm]mid1) -- ++(0,0.5);
    \node[text=customgray, align=center, inner sep=2pt, fill=none, above, draw=none] at ([yshift=10mm]mid1) {design solution\\\& refined input};

    \path (node2) -- (node3) coordinate[midway] (mid3);
    \draw[customgray] ([yshift=3mm]mid3) -- ++(0,0.5);
    \node[text=customgray, align=center, inner sep=2pt, fill=none, above, draw=none] at ([yshift=10mm]mid3) {realizable\\ set};

    \path (node3) -- (node4) coordinate[midway] (mid4);
    \draw[customgray] ([yshift=3mm]mid4) -- ++(0,0.5);
    \node[text=customgray, align=center, inner sep=2pt, fill=none, above, draw=none] at ([yshift=10mm]mid4) {objective\\function};

    \node[text=emphgray, align=right, inner sep=2pt, fill=none, above, draw=none] at ([yshift=9.5mm,xshift=34mm]mid4) {manual\\input};

    \path (node1) -- (node2) coordinate[midway] (mid12);
    \draw[customgray] ([yshift=-3mm]mid12) -- ++(0,-0.5);
    \node[text=customgray, align=center, inner sep=2pt, fill=none, below, draw=none] at ([yshift=-10mm]mid12) {reachability\\analysis};

    \path (node2) -- (node3) coordinate[midway] (mid34);
    \draw[customgray] ([yshift=-3mm]mid34) -- ++(0,-0.5);
    \node[text=customgray, align=center, inner sep=2pt, fill=none, below, draw=none] at ([yshift=-10mm]mid34) {CSP};

    \path (node4) -- (subcontracts) coordinate[midway] (mid46);
    \draw[customgray] ([yshift=-3mm]mid46) -- ++(0,-0.5);
    \node[text=customgray, align=center, inner sep=2pt, fill=none, below, draw=none] at ([yshift=-10mm]mid46) {optimization};

    \node[text=emphgray, align=left, inner sep=2pt, fill=none, below, draw=none] at ([yshift=-10mm, xshift=-29mm]mid12) {automated\\solvers};

\end{tikzpicture}
\caption{The overall correct-by-construction scheme for top-to-bottom RE. The operations on top require manual input systematically augmented by the automated outputs from the tools on the bottom.}
\label{fig10}
\end{figure}

The process initiates with the manual input of the design solution and the refined input set. First, conditions (\ref{cond1}) and (\ref{cond2}) are automatically checked to verify if the design solution refines the upper-level requirement comprehensively. If condition (\ref{cond1}) is not met, a new design solution is sought; similarly, if condition (\ref{cond2}) is not met, a new refined input set is manually selected. Once conditions (\ref{cond1}) and (\ref{cond2}) are satisfied, the lower bound of the subcontracts, indicated by condition (\ref{lobound}), is determined using the reachable sets identified in the reachability analysis.

Subsequently, the realizable sets in condition (\ref{upbound}) are input to automatically compute the upper bound of the subcontracts through constraint programming. Failing to find a feasible solution indicates conflicts within the realizable sets, implying supplier incompatibility. Similarly, if the contracted realizable sets do not encompass their corresponding reachable sets, some suppliers may be unable to provide the necessary devices for system operation.

Finally, the requirement parameters are recast into an optimization problem to determine the optimal subcontracts. The system automatically derives the constraints from conditions~(\ref{eq:constraints}). However, the objective function (\ref{optimization}) requires manual input based on the designer's specific goals, a reasonable expectation in the context of RE.

\paragraph{Top down vs. bottom up}
Firstly, our method is driven by the top-level requirement, ensuring that every lower-level component directly contributes to satisfying the overall system requirement. This contrasts with bottom-up approaches, such as those based on process algebras or component-based design, which typically start with predefined components and attempt to compose them to meet system-level goals. Our approach maintains traceability throughout the design process, linking each sub-requirement explicitly to higher-level objectives.

Secondly, we incorporate formal verification at each decomposition step. Unlike purely bottom-up approaches where compatibility issues may only become apparent during integration, our method allows designers to ensure that the proposed decomposition satisfies the higher-level requirement before proceeding further. This early validation is particularly valuable in complex systems where emergent properties may not be evident from individual component specifications alone.

Furthermore, our method supports an iterative refinement process guided by formal and \emph{designer-led} feedback. This allows for adjustments to be made early in the design process, driven by top-level requirements, rather than being constrained by existing component implementations. In contrast, bottom-up approaches often struggle to accommodate changes in system-level requirements, potentially requiring significant re-engineering of component interfaces and interactions.

While our approach begins with a top-down decomposition, it does not preclude the integration of existing components or bottom-up considerations. The formal contracts used in our method provide a rigorous interface between high-level requirements and low-level implementations. This allows designers to incorporate implementation constraints or existing components into the top-down design process, effectively bridging the gap between top-down design intentions and bottom-up implementation realities.

In practice, our method offers a more flexible and comprehensive approach to system design than purely bottom-up methods. By maintaining different levels of abstraction throughout the design process, it allows for the integration of both high-level design concepts and low-level implementation details. This can lead to more coherent system designs and potentially reduce integration issues that often plague bottom-up approaches.

In summary, while the initial decomposition in our method involves manual input from the designer, the overall approach maintains a top-down nature through its requirement-driven process, formal verification at each step, and ability to iteratively refine the design based on system-level requirements. This approach offers distinct advantages over traditional bottom-up methods, particularly in handling complex systems with evolving requirements that are capturable in continuous bounded sets.

\section{Cruise control requirements decomposition}
To illustrate the semi-automatic requirement decomposition process, we consider the design of a cruise control system. This system maintains the vehicle's speed within specific bounds under various operational conditions. The high-level requirement defines the speed parameters in natural language as follows: 

\vspace{1em}
\noindent
\emph{For any initial speed \(v(0)\) within \SIrange{23.0}{30.0}{\metre\per\second} and a reference speed \(v_{\text{ref}}\) within \SIrange{34.5}{35.5}{\metre\per\second}, the actual speed \(v(t)\) must remain within \SIrange{20}{40}{\metre\per\second} at all times during a \SI{100}{\second} simulated period. Furthermore, after \SI{20}{\second}, the speed should stabilize within \SIrange{33.5}{36.5}{\metre\per\second}.}
\vspace{1em}

This high-level requirement is deliberately defined with a degree of looseness to facilitate lower-level refinement and adaptation.

The contract to be refined, denoted as \(f^0\), is characterized by specific input and output sets that dictate the expected system behavior. For any \(v_{\text{ref}}\) and \(v(0)\) within the input set, the corresponding \(v(t)\) must adhere to the defined output set boundaries:

\begin{figure}[!t]
\centering
\begin{tikzpicture}[node distance=1cm, auto]
    \definecolor{wireColor}{RGB}{233,144,118}

    \node[draw] (g8) {$g^8$};
    \node[draw, right=of g8] (g4) {$g^4$};
    \node[draw, right=of g4] (g2) {$g^2$};
    \node[draw, right=of g2] (g1) {$g^1$};

    \node[draw, below=of g4] (g5) {$g^5$};
    \node[draw, above=of g4] (g3) {$g^3$};

    \node[draw, left=of g3] (g7) {$g^7$};
    \node[draw, left=of g7] (g6) {$g^6$};

    \draw[->, -latex] (g8) -- node[above] {$u(t)$} (g4);
    \draw[->, -latex] (g4) -- node[above] {$F(t)$} (g2);
    \draw[->, -latex] (g2) -- node[above] {$\dot{v}_t$} (g1);

    \node[above=-0.08cm of g8] {$f^8$};
    \node[above=-0.08cm of g4] {$f^4$};
    \node[above=-0.08cm of g2] {$f^2$};
    \node[above=-0.08cm of g1] {$f^1$};

    \node[above=-0.08cm of g5] {$f^5$};
    \node[above=-0.08cm of g3] {$f^3$};
    \node[above=-0.08cm of g7] {$f^7$};
    \node[above=-0.08cm of g6] {$f^6$};

    \draw[->, -latex] (g1.east) -- ++(1,0) node[above, midway, text=wireColor] (vt) {$v(t)$};

    \draw[->, latex-] (g1.south) -- ++(0,-.5) node[below, text=wireColor] {$v(0)$};

    \draw[->, latex-] (g8.south) -- ++(0,-.5) node[below, text=wireColor] {$v_\text{ref}$};

    \draw[->, -latex] (g5.east) -- (g2.south west) node[midway, right] {$Fa(t)$};

    \draw[->, -latex] (g3.east) -- (g2.north west) node[midway, right] {$Fr(t)$};

    \draw[->, -latex] (g6) -- node[above] {$\omega(t)$} (g7);

    \draw[->, -latex] (g7.east) -- node[right] {$T(t)$} (g4);

    \draw[->, -latex] ([xshift=.5cm]g1.east) |- ([yshift=-2.2cm]g1.east) -| (g5.south);

    \coordinate (midpoint) at ([yshift=-2.2cm]g1.east);
    \draw[->, -latex] (midpoint) -- ++(-7.5cm,0) |- (g8.west);
    \draw[->, -latex] (midpoint) -- ++(-7.5cm,0) |- (g6.west);

\end{tikzpicture}
\caption{A candidate design solution for a cruise control system (adapted from Aström and Murray~\citep{fbs}).}
\label{fig13}
\end{figure}
\[f^0 \colon v_{\text{ref}} \times v(0) \to v(t)\]
where 
\begin{align*}
 v_{\text{ref}} &= [34.5, 35.5]\; \si{\metre\per\second},\\
         v(0) &= [23.0, 30.0]\; \si{\metre\per\second}, \text{ and}\\
v(t) &= \begin{cases} [20, 40] \text{ m/s} & \text{for } t \in [0, 100] \text{ s} \\ [33.5, 36.5] \text{ m/s} & \text{for } t \in [20, 100] \text{ s} \end{cases}.
\end{align*}

By setting these high-level requirements and defining the input and output sets for \(f^0\), we lay the groundwork for the subsequent stages of system design and refinement, ensuring that the final implementation aligns with the desired performance criteria and operational constraints.

The design solution for the cruise control system (\cref{fig13}) demonstrates the semi-automatic requirement decomposition process. The objective of this section is not to develop the optimal cruise control system but to use it as a case study for explaining the decomposition process---often, we have to sacrifice optimal behavior for safety. The system is articulated through eight functions ($g^1, g^2, \ldots, g^8$) (\cref{tab2}) and comprises eight variables: $$(v, \dot{v}, Fr, F, Fa, T, \omega, u)^\top = (x_1, x_2, \dots, x_8)^\top.$$ Key uncertain parameters include the total mass ($m$) and the maximal engine speed ($\omega_m$). The input sets are extended to accommodate the design solution's refinement of the high-level contract $f^0$, with $v_\text{ref}$ in the range of \SIrange{34.0}{36.0}{\meter\per\second} and $v(0)$ within \SIrange{22.0}{30.0}{\meter\per\second}.

\begin{figure}[!t]
\centering
\includegraphics[width=.7\columnwidth]{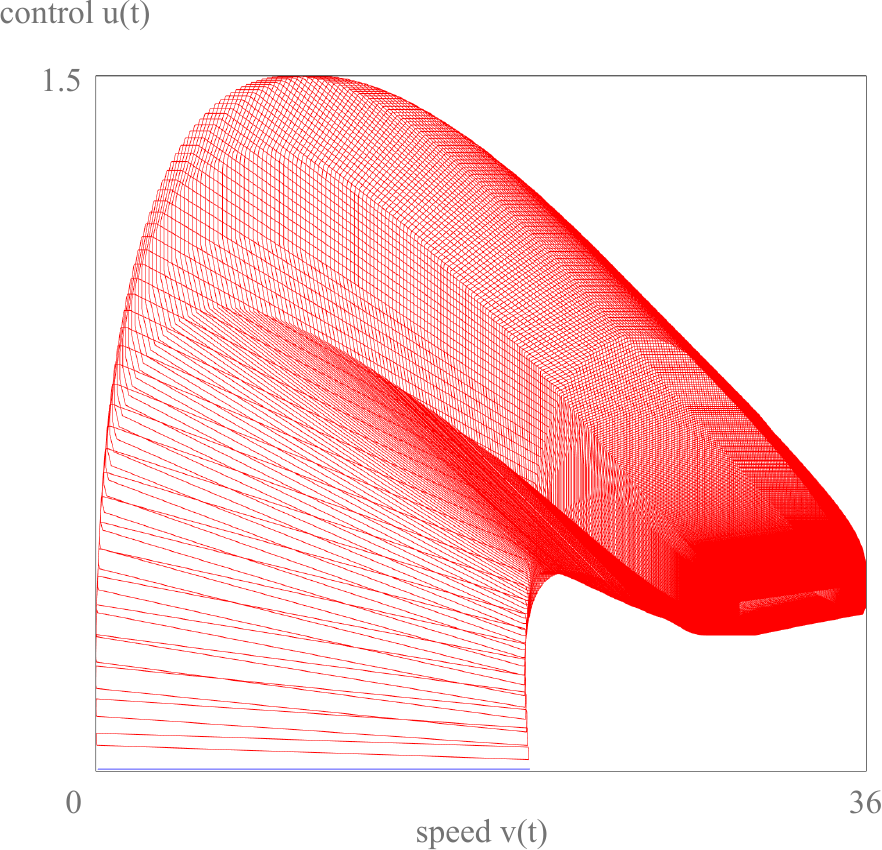}
\caption{Evolution of the reachable sets of  $(v(t), u(t))^\top$ during $t\in [0, 100]$.}
\label{fig^14}
\end{figure}

\begin{figure}[!t]
\centering
\includegraphics[width=.7\columnwidth]{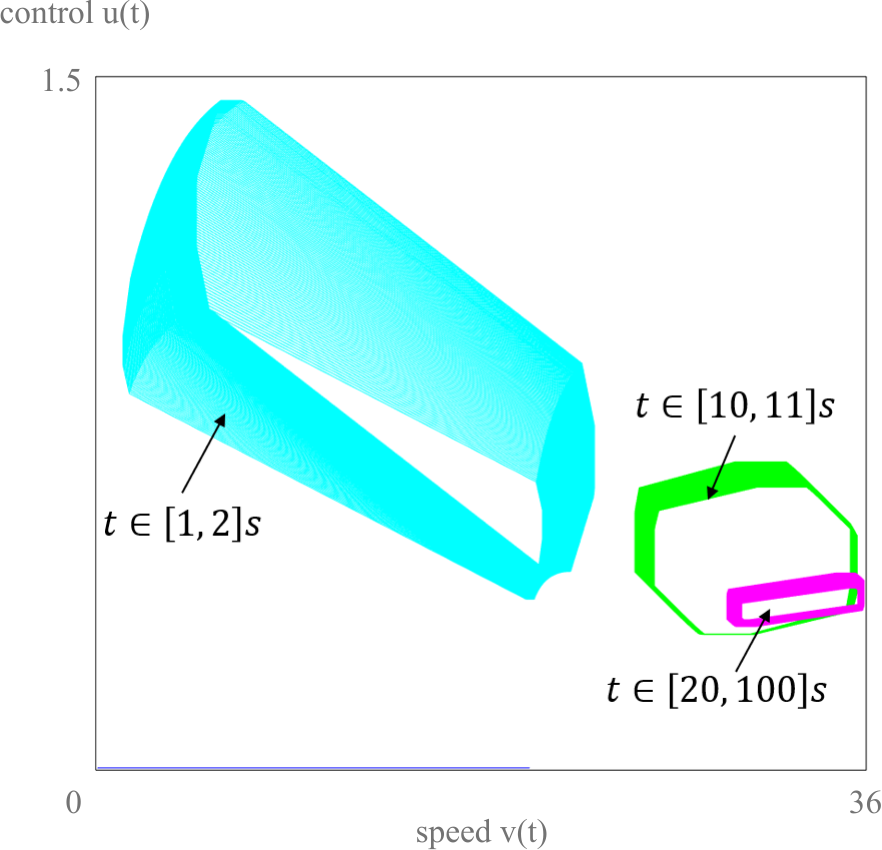}
\caption{Discrete sampling of  reachable sets $(v(t), u(t))^\top$ during $t\in [0, 100]$.}
\label{fig^15}
\end{figure}

The reachable sets for the system variables, indicative of the range of possible states the system can attain over time, are computed using CORA \citep{cora}. The evolution of the system state over the simulation period is characterized by a transition from a broad initial state to a progressively more focused and stable region. Initially, the system state starts from a specific set of conditions and evolves, as demonstrated by the progression and narrowing of the reachable sets (\cref{fig^14,fig^15}).

Specifically, the system begins with a wide range of possible states, which is then gradually confined to a smaller region, reflecting the system's convergence to a stable state. This dynamical behavior is evidenced by a continuous plot showing how the system's state converges over time, beginning with a broad spread and then moving towards a more concentrated area, indicative of system stabilization (\cref{fig^14}). Additionally, discrete snapshots of the system's state provide insight into its temporal progression, showing a narrowing of the possible states as time advances (\cref{fig^15}). The detailed bounds of the reachable sets for each variable provide insight into the operational limits and capabilities of the system (\cref{tab3}).

\begin{table*}[!
htbp
]
\centering
\caption{The functions and parameters of the design solution.}
\label{tab2}
\renewcommand{\arraystretch}{3}
\mbox{}\clap{
\begin{tabular}{|l|l|p{8.5cm}|}
\hline
$g^i$ &  Function &  Explanation\\ \hline
 $x_1=g^1(x_2)$& $v(t)=\int_{0}^{t} \dot{v} dt+v(0)$ & The definition of speed change \\ \hline
 $x_2=g^2(x_3, x_4,x_5,m)$& $\dot{v(t)}=(F(t)-Fr-Fa(t))/m$  & The driving force $F(t)$, the aerodynamic drag $Fa(t)$ and the tyre friction $Fr$ are the force considered.  \\ \hline
 $x_3=g^3(m)$&$Fr=mgC_r$  & The tyre friction, where $C_R$ is coefficient for the tyre friction. \\ \hline
 $x_4=g^4(x_6, x_8)$& $F(t)=\alpha*u(t)T(t)$ & The driving force, where $\alpha$ is the gear rate, $u(t)$ is the control input, $T(t)$ is the torque.\\ \hline
 $x_5=g^5(x_1)$& $Fa(t)=\frac{1}{2}\rho C_dAv(t)^2$  &  The aerodynamic drag, where $\rho$ is the air density, $C_d$ is the shape-dependent aerodynamic drag coefficient and $A$ is the frontal area of the car.  \\ \hline
 $x_6=g^6(x_7,
\omega_m)$&$T(t)=T_m(1-\beta(\frac{\omega(t)}{\omega_m}-1)^2)$  & The torque is correlated to the engine speed $\omega(t)$ (RPM) and the maximum torque $T_m$ is obtained at engine speed $\omega_m$. \\ \hline
 $x_7=g^7(x_1)$& $\omega(t)=\alpha*v(t)$  &  The engine speed (RPM) is correlated with the car speed. \\ \hline
 $x_8=g^8(x_1)$& $u(t)=p(v_\text{ref}-v(t))+i\int_{0}^{t}(v_\text{ref}-v(t)) dt$ &  The PI controller. \\ \hline
   \multicolumn{3}{|p{15.5cm}|}{$x_1=v, x_2=\dot{v}, x_3=F_r, x_4=F, x_5=F_a, x_6=T, x_7=\omega, x_8=u$} \\ \hline
 \multicolumn{3}{|p{17cm}|}{$m\in [900,1100], \omega_m\in [365,450], v_\text{ref}\in[34.0, 36.0], v(0)\in [22.0,30.0], \alpha=10, g=9.8, \newline C_r=0.01, C_d=0.32, \rho=1.3, A=2.4, \beta=0.4, T_m=200, p=0.1, i=0.5$} \\ \hline
\end{tabular}%
}
\end{table*}

Firstly, the torque \(T(t)\) is constrained within \(\SIrange{179.939}{200.000}{}\), adhering to the maximum torque limit \(T_m = \SI{200}{}\). Similarly, the engine speed \(\omega(t)\) is bound within \(\SIrange{219.665}{362.297}{}\), respecting the maximum engine speed range of \(\SIrange{365}{450}{}\). These constraints ensure that the system's operational parameters remain within the physical capabilities of the vehicle.

Secondly, the design solution's output aligns with the upper-level contract's stipulated requirements. Conditions~(\ref{cond1}) and (\ref{cond2}) are fulfilled, indicating that the design solution effectively refines the high-level contract. Consequently, the reachable sets serve as the lower bounds for the subcontracts, satisfying condition (\ref{lobound}). Specifically, for \(t \in \SIrange{0}{100}{\second}\), the vehicle's speed \(v(t)\) is maintained within \(\SIrange{21.966}{36.230}{\meter\per\second}\), comfortably within the broader speed range of \(\SIrange{20.0}{40.0}{\meter\per\second}\). For the latter part of the trajectory, \(t \in \SIrange{20}{100}{\second}\), the speed \(v(t)\) narrows down to \(\SIrange{33.651}{36.192}{\meter\per\second}\), aligning with the more stringent requirement of \(\SIrange{33.5}{36.5}{\meter\per\second}\). 

\begin{table}[!t]
\centering
\caption{The reachable sets during $t\in [0, 100]$ seconds.}
\label{tab3}
\renewcommand{\arraystretch}{3}
\begin{tabular}{|l|l|}
\hline
 Variable& Reachable set \\ \hline
 \multirow{2}{*}{$x_1=v(t)$}& $[21.96649,36.22973]$  \\ \cline{2-2} 
 &  $[33.65083,36.19185]$ for $t > \SI{20}{\second}$ \\ \hline
 $x_2=\dot{v(t)}$&$[-0.62937,2.89036]$  \\ \hline
 $x_3=Fr(t)$& $[88.20000,107.80000]$ \\ \hline
 $x_4=F(t)$ &$[-9.33497,2997.78280]$  \\ \hline
 $x_5=Fa(t)$&  $[240.87740,655.24649]$\\ \hline
 $x_6=T(t)$& $[179.93903,200.00000]$ \\ \hline
 $x_7=\omega(t)$&  $[219.66494,362.29727]$\\ \hline
 $x_8=u(t)$ &$[-0.00472,1.50660]$  \\ \hline
\end{tabular}
\end{table}

\subsection{Realizable sets}
The formulation of the design solution into a CSP yields three possible outcomes.
\begin{enumerate}
    \item Successful contraction of the original sets encompassing the lower bounds, ensuring that all necessary operational parameters are included.
    \item Successful contraction of the original sets but excluding some lower bounds might indicate the need for further refinement or adjustment in the system's parameters or constraints.
    \item Failure to contract, suggesting internal conflicts among the original sets. This outcome reveals inconsistencies or infeasibilities within the system design that require resolution.
\end{enumerate}

Various test cases are presented (\cref{tab4}) to illustrate different possible outcomes from the CSP formulation, assuming consistency in the original realizable sets across variables for generality. Test 1 demonstrates the successful contraction of the original sets, with the resulting sets exceeding the lower bounds and qualifying as upper bounds for the subcontracts.

In Test 2, the torque's realizable set \(T(t)\) is modified to \(\SIrange{0}{180}{}\). This adjustment results in contracted sets of \(v(t)\), \(\omega(t)\), \(T(t)\), and \(Fa(t)\) that do not cover the respective lower bounds. This suggests a potential need for a device capable of providing a broader torque range. 

\begin{table}[!t]
\centering
\caption{Test cases for the realizable sets.}
\label{tab4}
\renewcommand{\arraystretch}{2.5}
\begin{tabular}{|l|l|l|l|}
\hline
 & \textbf{Test 1} &\textbf{Test 2}  & \textbf{Test 3}  \\ \hline
 Variable & \multicolumn{3}{c|}{Original realizable sets} \\ \hline
$x_1=v(t)$ &  [0,50]  & [0,50] & [0,50] \\ \hline
$x_2=\dot{v(t)}$ &  [-1.5,3] &[-1.5,3] &  [-1.5,3]  \\ \hline
$x_3=Fr(t)$ &  [70,120] & [70,120] &   [70,120] \\ \hline
$x_4=F(t)$ &  [-250,3500] & [-250,3500] &  [-250,3500]  \\ \hline
$x_5=Fa(t)$& [0,1000] & [0,1000] & \textbf{[600, 1000]}  \\ \hline
$x_6=T(t)$&  [0,250] & \textbf{[0,180]} &[0,180] \\ \hline
$x_7=\omega(t)$ &  [0,450] & [0,450] & [0,450]  \\ \hline
$x_8=u(t)$ & [-0.5,2] & [-0.5,2] &[-0.5,2]  \\ \hline
Variable & \multicolumn{3}{c|}{Contracted realizable sets} \\ \hline
$x_1=v(t)$ &  [0, 45.000]  & \textbf{[0, 33.000]} & \multirow{8}{*}{Empty} \\ \cline{1-3}
$x_2=\dot{v(t)}$ &  [-1.5, 3.0] & [-1.0, 3.0]  &  \\ \cline{1-3}
$x_3=Fr(t)$  & [88.2,107.8]  & [88.2, 107.8]  &  \\ \cline{1-3}
$x_4=F(t)$ & [-250.0, 3500.0] & [-250,3500.0]   &  \\ \cline{1-3}
$x_5=Fa(t)$ &  [0,1000.0]  &\textbf{[0, 543.6]}  &  \\ \cline{1-3}
$x_6=T(t)$ &  [120.0, 220.0] & \textbf{[120.0, 180.0]}  &  \\ \cline{1-3}
$x_7=\omega(t)$&  [0,450.0]  &\textbf{[0,330.0]}   &  \\ \cline{1-3}
$x_8=u(t)$&  [-0.208, 2.0] & [-0.208, 2.0]  &  \\ \hline
\end{tabular}%
\end{table}

Test 3 modifies the aerodynamic drag \(Fa(t)\) to a range of \(\SIrange{600}{1000}{}\) compared to Test 2. This change yields no feasible contracted realizable sets, indicating a significant internal conflict. Notably, the contracted set of \(Fa(t)\) in Test 2 is \(\SIrange{0}{543.6}{}\), which has no overlap with the original set of \(Fa(t)\) in Test 3 (\(\SIrange{600}{1000}{}\)), leading to the conflicts observed.

The contracted realizable sets from Test 1 are adopted as the upper bounds for the subcontracts, satisfying condition (\ref{upbound}).

Formulating the design solution into a CSP is a formal approach to refining and validating system requirements and designs. It categorically addresses the system's feasibility through three possible outcomes, each offering specific insights into its design quality. Successful contraction ensures all operational parameters are included, indicating a robust design. Contraction excluding some lower bounds signals a need for further refinement, prompting engineers to reassess and tweak system parameters or constraints. Failure to contract points out internal conflicts, urging a thorough reevaluation of the system design to resolve inconsistencies or infeasibilities. The test cases exemplify how different scenarios affect the design's validity, guiding engineers to make informed decisions to enhance system performance and reliability. This systematic approach in CSP enables a clear, step-by-step refinement process, ensuring that the system meets its intended requirements and operates effectively within its defined parameters.

\subsection{Trade-off study: optimal subcontracts}
We formulate the optimization problem to determine the optimal subcontracts for the cruise control example (\cref{tab5}). The decision variables, \(\underline{X_i}\) and \(\overline{X_i}\), represent the lower and upper bounds of subcontract \(X_i\) for \(i = 1, 2, \dots, 8\). The objective function is tailored to the designer's preference and is constructed following condition (\ref{optimization}). The parameters required to calculate \(h_i^k\) are provided in \cref{tab6}.

\begin{table}[!t]
\centering
\caption{Optimization problem formulation for~achieving~optimal~subcontracts.}
\label{tab5}
\renewcommand{\arraystretch}{3}
\begin{tabular}{|l|l|}
\hline
 \makecell[l]{Decision\\variables} &  $\underline{X_i}$ and $\overline{X_i}$ where $i=1,2,...,8$\\ \hline
\makecell[l]{Objective\\function}& $\textit{minimize} \sum_{k=1}^8\sum_{i=1}^8 h_i^k$ \\ \hline
 Constraints & 
 \makecell[l]{
 \\[-0.5em]
  $\overline{RA_{x_i}}\leq \overline{X_i}\leq \overline{RE_{x_i}}$\\[0.4em]
 $\underline{RE_{x_i}}\leq \underline{X_i}\leq \underline{RA_{x_i}}$\\[0.4em]
$\underline{X_1}\geq 20, \overline{X_1}\leq 40$\\[0.4em]
$\underline{X_7}\leq g^7(\underline{X_1})$\\[0.4em] $\overline{X_7}\geq g^7(\overline{X_1})$\\[0.4em]
$\underline{X_6}\leq g^6(\underline{X_7}, \overline{\omega_m})$\\[0.4em] $\overline{X_6}\geq g^6(\overline{X_7},\underline{\omega_m})$\\[0.4em]
$\underline{X_5}\leq g^5(\underline{X_1})$\\[0.4em] $\overline{X_5}\geq g^5(\overline{X_1})$\\[0.4em]
$\underline{X_4}\leq g^4(\overline{X_6},\underline{X_8})$\\[0.4em] 
$\overline{X_4}\geq g^4(\overline{X_6},\overline{X_8})$\\[0.4em]
$\underline{X_2}\leq g^2(\underline{X_4}, \overline{X_3},\overline{X_5}, \overline{m})$\\[0.4em] $\overline{X_2}\geq g^2(\overline{X_4}, \underline{X_3},\underline{X_5}, \underline{m})$\\[0.4em]
    $\underline{\omega_m}=365, \overline{\omega_m}=450, \underline{m}=900, \overline{m}=1100$\\[0.4em]}
  \\ \hline
\end{tabular}
\end{table}

The constraints for the optimization are established according to (\ref{optimization}), where \(x = (x_1, x_2, \ldots, x_8)^\top\). The reachable sets \(RA_{x_i}\) and realizable sets \(RE_{x_i}\) are documented in \cref{tab6}, and the functions \(g^i\) are outlined in \cref{tab2}.

The optimal subcontracts from the formulation are depicted in \cref{tab7}. The given contract at level 0 is refined by the specifications in the second row, while the decomposed requirements, or the subcontracts, are presented in the third row (\cref{tab7}).

\begin{table}[!t]
\caption{The parameters for the trade-off study.}
\label{tab6}
\renewcommand{\arraystretch}{3}
\begin{tabular}{|l|l|l|l|}
\hline
$X_i$ &  $[\underline{RA_{x_i}},\overline{RA_{x_i}}]$ & $[\underline{RE_{x_i}},\overline{RE_{x_i}}]$&  Difficulty coefficients\\ \hline
 $X_1$& $[22.00,36.23]$ &$[0, 45]$ & \makecell{$a_1^1=0.9, a_1^5=0.5$,\\ $a_1^7=0.1,a_1^8=0.8$}  \\ \hline
 $X_2$&  $[-0.63,2.89]$  & $[-1.5, 3.0]$  & $a_2^1=0.6, a_2^2=0.6$ \\ \hline
 $X_3$&  $[88.20,107.80]$ & $[88.2,107.8]$  & $a_3^3=0.5, a_3^2=0.4$ \\ \hline
 $X_4$&  $[-9.36,2997.78]$  &$[-250.0, 3500.0]$  & $a_4^4=0.3, a_4^2=0.8$ \\ \hline
 $X_5$&  $[240.88,655.25]$ &  $[0,1000.0]$  & $a_5^5=0.4, a_5^2=0.7$ \\ \hline
 $X_6$&  $[179.94,200.00]$ & $[120.0, 220.0]$  &  $a_6^6=0.1, a_2^2=0.9$\\ \hline
 $X_7$&  $[219.67,362.30]$ &  $[0,450.0]$  &  $a_7^7=0.5, a_7^6=0.6$\\ \hline
 $X_8$&  $[-0.005,1.51]$ & $[-0.29, 2.0]$  & $a_8^8=0.9, a_8^4=0.9$ \\ \hline
\end{tabular}
\end{table}

Control input \(u\) is included at level 0, and notably, \(f^1\) and \(f^8\) are absent from the level 1 subcontracts. This is because speed \(v\) and control input \(u\) are system-wide properties and cannot be relegated to lower levels. Specifically, for \(f^1\), reasoning about \(v(t)\) from \(\dot{v(t)}\) requires knowledge of the entire system's behavior over time, which is determined by the composition of the whole system. However, at level 0, \(v(t)\) can be contextualized with the inputs \(v(0)\) and \(v_\text{ref}\). This relationship implies that \(v(t)\) is inherently associated with the highest level of system abstraction and must be ensured at level~0. Accordingly, \(g^1\) and \(g^8\) do not form part of the optimization constraints (\cref{tab5}).

By defining explicit upper and lower bounds for each subcontract, the framework precisely delineates each component's operational space, directly linking the technical specifications to the system's overall objectives. This precision in specifying requirements ensures that each subsystem or component meets its individual objectives and aligns with the overarching system goals, enhancing the coherence and efficiency of the system as a whole. The optimization process iteratively refines these subcontracts, allowing for a  calibration of requirements responsive to internal system constraints and external performance criteria. 

\subsection{Discussion}
This paper has introduced a formalized approach for semi-automatic requirement decomposition based on CBD. This research augments a top-down derivation of sub-requirements common in systems engineering with formal methods typically used in bottom-up compositional verification, offering a more systematic and intuitive way to bridge high-level system goals with detailed technical specifications. The formalism developed ensures mathematical correctness and provides a structured method for refining system requirements, further enhanced through integrating formal methods in reachability analysis and CSP.

The results from applying this methodology, notably demonstrated through the cruise control system case study, underscore its practicality. The approach not only streamlines the process of requirement decomposition but also enhances the clarity and reliability of the resulting system specifications. However, the critical insight gleaned from this work is identifying a significant gap in traditional RE practices: the need for a systematic, top-down approach that can effectively handle modern systems' complexity and multi-level nature. The proposed approach fills this gap by offering a structured and formalized method that aligns with the nuanced demands of system design.

Furthermore, the case study validates the possible depth of analysis and refinement. Yet, it also brings to light the challenges and limitations associated with modeling requirements as continuous bounded sets, especially for systems that might be more accurately represented as transition systems dealing with discrete traces. This recognition opens further research and development avenues, emphasizing the need to extend and adapt the methodology to a broader range of system types and complexities. The ultimate aim is to create a robust toolkit for RE that can adapt to the evolving system design landscape, thereby enhancing system development's efficiency, reliability, and traceability in the RE field.

Despite the current methodology's emphasis on continuous bounded sets, its core principles and formal approach retain their applicability to transition systems. The transition system, a model dealing with discrete states and events, requires a nuanced treatment of requirements and sub-requirements that are inherently discrete and often complex. While the current approach may necessitate adjustments and extensions to fully encapsulate these systems' discrete nature, the foundational goal of ensuring a mathematically correct, top-down decomposition of requirements remains unchanged. By incorporating concepts and tools suited to discrete systems, such as state-transition matrices, automata theory, or model-checking algorithms, the methodology can be adapted to address the unique challenges and characteristics of transition systems. This adaptation would further demonstrate the methodology's potential as a comprehensive toolkit in RE, capable of accommodating the diverse modeling needs of modern complex systems, thereby maintaining the rigor and structured approach central to enhancing system design and development processes.

By recognizing the limitations of current practices, particularly in handling diverse and complex systems, this research advocates for a more adaptable, robust approach. Incorporating formal methods, such as reachability analysis and CSP, into RE enhances precision and rigor, ensuring that each component's specifications are theoretically sound and practically viable. The emphasis on a structured, top-down approach reflects an understanding of the relationships and dependencies within modern systems, advocating for a strategy that is both comprehensive and flexible. As this methodology continues to evolve and adapt, it can provide practitioners with a powerful tool to navigate the complexities of system design with greater confidence and efficacy. Pursuing a more adaptable, inclusive approach is not just an academic exercise but a strategic imperative to keep pace with the rapid technological advancements and the increasing sophistication of system designs, particularly in cyber-physical systems.

\begin{table}[!t]
\caption{The correct-by-construction optimal subcontracts.}
\label{tab7}
\renewcommand{\arraystretch}{3}
\begin{tabular}{|l|c|l|l|}
\hline 
\multicolumn{2}{|l|}{} &Input  & Output \\ \hline
\multicolumn{2}{|l|}{\makecell[l]{Level 0\\ contract}} &  \makecell[l]{$v_\text{ref}\in[34.5,35.5]$\\$v(0)\in [22.0,30.0]$}& $v\in [20, 40]$\\ \hline
\multicolumn{2}{|l|}{\makecell[l]{Level 0\\ refinement}} & \makecell[l]{$v_\text{ref}\in[34.0, 36.0]$\\$v(0)\in [22.0,30.0]$} & \makecell[l]{$v\in [21.8, 38.4]$\\ $u\in [-0.1, 1.511]$} \\ \hline
\multirow{6}{*}{\makecell[l]{Level 1\\subcontract}} & $f^2$ & \makecell[l]{$Fr\in[88.2,107.8]$\\$F\in [-159.1,3024.0]$\\$F_a\in [237.6,827.6]$} & $\dot{v}\in [-1.1, 3]$ \\ \cline{2-4} 
 & $f^3$ &  NA& $Fr\in[88.2,107.8]$ \\ \cline{2-4} 
 &  $f^4$& \makecell[l]{$T\in[150, 200.1]$\\$u\in [-0.1, 1.511]$} & $F\in [-159.1,3024.0]$ \\ \cline{2-4} 
 &  $f^5$& $v\in [21.8, 38.4]$ &  $F_a\in [237.6,827.6]$\\ \cline{2-4} 
 &  $f^6$& $\omega\in [109.8,406.1]$ &  $T\in[150, 200.1]$\\ \cline{2-4} 
 &  $f^7$& $v\in [21.8, 38.4]$ & $\omega\in [109.8,406.1]$ \\ \hline
\end{tabular}
\end{table}

\paragraph{A note on requirement capture and computational tools for practitioners}

This case study was conducted using MATLAB to perform the computational aspects of the methodology. While natural language requirements were manually translated within MATLAB for illustrative purposes, the practicality of the approach can be enhanced by leveraging existing toolchains.
For \emph{formal requirements capture}, structured methods such as the Easy Approach to Requirements Syntax (EARS)~\citep{DBLP:conf/re/MavinWHN09} are widely used in RE and could serve as a valuable intermediary layer between natural language specifications and contract theory as a \emph{requirements formalism}.
\emph{Reachability analysis} was performed using CORA, a common choice for continuous systems. However, alternative tools exist for different use cases. For instance, SpaceEx~\citep{DBLP:conf/cav/FrehseGDCRLRGDM11} is well-suited for hybrid systems, addressing both continuous and discrete behaviors.
\emph{Constraint solving and optimization}, can be effectively handled by established tools such as Google OR-Tools or IBM CPLEX, depending on the specific problem formulation and complexity.
To facilitate practical implementation, \emph{integrating the development environment} is crucial. Options include expanding the MATLAB environment with custom requirements capture functionality, or developing integrated solutions using platforms like Eclipse and Jupyter notebooks. These integrated environments would streamline the workflow from requirement specification to analysis and optimization.

\section{Related Work}
One of the issues in introducing design errors in embedded systems is due to a lack of formalism in translating system requirements into subsystem requirements and subsequently into code \citep{methodology}. In particular, Moore supported this observation by highlighting that existing formal techniques predominantly focus on the bottom-up composition of component-level specifications into system-level specifications rather than on a top-down derivation of component requirements from higher-level system requirements \citep{specification}. Despite the passage of time, the literature continues to be skewed towards bottom-up formalisms and component-based composition, with scant attention to top-down decomposition. This paper aims to address this gap by concentrating on top-down requirement decomposition.

Industry standards delineate system and software requirement document structure \citep{ieeesys}. Arora et al. introduced the concept of requirement boilerplates to confine the syntax of requirement sentences to predefined linguistic structures, significantly reducing ambiguity and enhancing the feasibility of automated analysis \citep{boilerplates}. Mahendra and Ghazarian conducted extensive studies on requirements patterns to aid practitioners in identifying, analyzing, and structuring software system requirements~\citep{patterns}. Further, Reinkemeier et al. developed a pattern-based requirement specification language specifically targeting timing requirements in the automotive domain \citep{timing}. Project CESAR adopts a formalism that translates requirements from natural language into boilerplates and pattern-based requirements for automated analysis \citep{cesar}. While these formalisms contribute to reducing ambiguity in requirement expression, they do not incorporate formal approaches for decomposition.

In component-based composition, the architecture analysis \& design language~(AADL) is recognized as a leading architecture description language, offering formal modeling concepts for describing and analyzing system architecture through components and their interactions \citep{AADL}. Bozzano et al. developed a model-checker for the functional verification and safety analysis of AADL-based models \citep{mcaadl}. Hu et al. presented a methodology for translating AADL into a timed abstract state machine to verify both functional and nonfunctional properties of AADL models \citep{aadltrans}. Johnsen et al. introduced a formal verification technique for automated fault avoidance in systems specified in AADL \citep{automated}. Despite AADL's support for step-wise refinement and architectural verification, it does not facilitate the top-down automatic generation of sub-requirements. Other frameworks such as EAST-ADL, Rodin, and FDR contribute to the landscape \citep{eastadl,rodin,fdr}.

Recently, CBD has emerged as an increasingly influential paradigm \citep{cbd2012}. One of its aims is to guide the initial specification process in a precise manner and assess quality-related properties using computational tools \citep{cbd20121}. The theory behind CBD has seen application across various domains. For instance, contract formalism has been utilized in the control design of hybrid systems, virtual testing, architecture design, and the development of property-based proof systems~\citep{taming,2008contract,2011using,2012property}. Tools designed for CBD, like AGREE, MICA, and OCRA, have been instrumental in advancing the RE field \citep{agree,mica,ocra}.

Assume-guarantee reasoning has been a powerful technique in CBD for decomposing verification tasks of complex systems into subtasks on individual components. \citet{10.1145/3243216} present an assume-guarantee contract framework for cyber-physical system design under probabilistic requirements specified in Stochastic Signal Temporal Logic (StSTL). \citet{10.1007/978-3-540-71209-1_21} define and study the co-synthesis problem, formulating it as an assume-guarantee synthesis problem to synthesize strategies that guarantee desired specifications. \citet{10.1145/509705.509707} present an assume-guarantee rule for checking simulation relations between state transition systems, enabling decomposition of simulation checks. \citet{10.1007/3-540-45351-2_24} introduce a model for hierarchical system design supporting an assume-guarantee principle for mixed parallel-serial contexts, suitable for modeling embedded systems interacting with real-world environments. \citet{10.1007/3-540-48153-2_30} extend circular compositional reasoning to handle liveness properties using a new rule implemented in the SMV proof assistant.

Nonetheless, these methods and tools are predominantly geared towards verification (bottom-up composition) and do not provide a methodology to generate lower-level subcontracts from upper-level contracts precisely and systematically. As highlighted in the contracts literature, transitioning from high-level contracts to an architecture of sub-contracts remains a challenging task often done manually \citep{2015contracts}. This paper responds to this challenge through semi-automated systems engineering holistically as opposed to other works that focus on synthesizing contracts for a specific instance of component \citep{santos:2019,dragomir:2017}---which we could incorporate as individual improvements when relevant within the design steps (\cref{fig10}).

\section{Conclusion}
In this paper, we have developed a formalized approach for semi-automatic requirement decomposition explicitly addressing a critical gap in RE. Our framework introduces a structured perspective in system design by shifting focus from the prevalent bottom-up compositional verification to a systematic top-down derivation of sub-requirements. The formalism established for requirement definition and decomposition ensures  correctness and facilitates a systematic refinement of system requirements. Integrating formal methods such as reachability analysis and CSP significantly enhances the automation and precision of the process, aligning with the practical needs and complexities of modern RE. Correct-by-construction commits to continuous refinement, ensuring systems are built to meet current specifications and adapt to future changes and challenges.

We must point out that this correct-by-construction decomposition does not work for any arbitrary requirement but only for ones we can quantify---otherwise understood as \emph{physical} requirements. However, the central insight from this work is that a structured, formal approach to top-down requirement decomposition for physical requirements bridges the gap between high-level system goals and detailed technical specifications. This not only aids in maintaining the integrity and traceability of system requirements and provides a precise roadmap for system architects and engineers. The case study on a cruise control system validates the approach, demonstrating its potential to simplify and clarify the RE process. As systems grow increasingly complex and integrated, the correct-by-construction offers a valuable tool for navigating requirement specification and decomposition, promising to enhance the efficiency and reliability of system design in RE by integrating formal methods early in the system's lifecycle as a form of ``what if'' analysis.

\bibliography{manuscript}
\end{document}